\documentclass[runningheads]{llncs}

\usepackage{multirow}
\usepackage{amsmath,amssymb,amsfonts}
\usepackage{algorithmic}
\usepackage{hyperref}
\usepackage{xurl}
\usepackage{xcolor}
\usepackage{enumitem}
\usepackage[T1]{fontenc}
\usepackage{orcidlink}
\usepackage{graphicx}
\usepackage{subcaption}

\makeatletter
\newcommand{\printfnsymbol}[1]{%
  \textsuperscript{\@fnsymbol{#1}}%
}
\makeatother

\usepackage{color}

\usepackage{xspace}

\begin{document}

\title{Investigating Mixture of Experts in Dense Retrieval}

\titlerunning{Investigating MoE in Dense Retrieval}

\author{Effrosyni Sokli\inst{1}\orcidlink{0009-0003-5388-2385} \and
Pranav Kasela\inst{1}\orcidlink{0000-0003-0972-2424} \and Georgios Peikos\inst{1}\orcidlink{0000-0002-2862-8209} \and
Gabriella Pasi\inst{1}\orcidlink{0000-0002-6080-8170}}

\authorrunning{Sokli et al.}

\institute{\email{\{effrosyni.sokli, pranav.kasela, georgios.peikos, gabriella.pasi\}@unimib.it} \vspace{4mm}\\ Department of Informatics, Systems and Communication, DISCo \\ University of Milano-Bicocca, Milan, Italy}

\maketitle

\begin{abstract}
While Dense Retrieval Models (DRMs) have advanced Information Retrieval (IR), one limitation of these neural models is their narrow generalizability and robustness.
To cope with this issue, one can leverage the Mixture-of-Experts (MoE) architecture. 
While previous IR studies have incorporated MoE architectures within the Transformer layers of DRMs, our work investigates an architecture that integrates a single MoE block (\texttt{SB-MoE}) after the output of the final Transformer layer.
Our empirical evaluation investigates how \texttt{SB-MoE} compares, in terms of retrieval effectiveness, to standard fine-tuning.
In detail, we fine-tune three DRMs (TinyBERT, BERT, and Contriever) across four benchmark collections with and without adding the MoE block.
Moreover, since MoE showcases performance variations with respect to its parameters (i.e., the number of experts), we conduct additional experiments to investigate this aspect further.
The findings show the effectiveness of \texttt{SB-MoE} especially for DRMs with a low number of parameters (i.e., TinyBERT), as it consistently outperforms the fine-tuned underlying model on all four benchmarks.
For DRMs with a higher number of parameters (i.e., BERT and Contriever), \texttt{SB-MoE} requires larger numbers of training samples to yield better retrieval performance.

\end{abstract}

\section{Introduction}
Neural Information Retrieval (IR) models, including several Dense Retrieval Models (DRMs), have shown a potential to enhance retrieval performance compared to sparse lexicon-based models such as BM25 \cite{robertson1995okapi}.
DRMs can be trained to capture the semantic context of queries and documents \cite{mitra2018introduction}. 
Nonetheless, their training requires large labeled datasets and there is a trade-off between generalizability and task-specific performance since DRMs often struggle to robustly adapt to different tasks or domains without additional fine-tuning.

In this paper, we investigate how an enhanced bi-encoder DRM architecture leveraging Mixture-of-Experts (MoE) \cite{jacobs1991adaptive} performs compared to the original underlying model across various dense retrieval scenarios.
Since MoE consists of expert sub-networks trained in an unsupervised manner, we explore if its addition to DRMs allows each expert to capture meaningful insights (e.g.,  topicality, text complexity), and therefore improve the final outputted embedding representation of the underlying model.
While previous works in IR have integrated MoE within each Transformer layer of the original model \cite{guo2024came,shen2024mixtureofexperts}, we apply a single MoE block (\texttt{SB-MoE}) on the output embeddings of the final Transformer layer of the underlying DRM. 
\texttt{SB-MoE} consists of multiple pairs of feed-forward networks (FFNs), where each pair acts as a unique expert.
The selection of the experts to be used each time is conducted by a gating function, which is a neural network trained in an \textit{unsupervised} manner to automatically assign an importance weight to every expert and select the optimal outcome based on a chosen pooling strategy.
The weights indicate how relevant each expert is to the current query or document input representation and determine their impact on the final prediction.
Hence, \texttt{SB-MoE} optimizes each expert and dynamically selects and fuses their outputs to tailor its predictions to the input data, which are the query and document representations outputted by the underlying model.
We apply \texttt{SB-MoE} to two datasets of the BEIR collection \cite{thakur2021beir}, and two datasets of the Multi-Domain Benchmark proposed by Bassani et al. \cite{bassani2022multidomain}, to conduct an empirical investigation on the usefulness of MoE enhancing DRMs on a query or document level for open-domain Q\&A, and domain-specific search.

This work has the following contributions: 
(1) we introduce a modular MoE architecture, \texttt{SB-MoE}, which takes as input the query and document embedding outputted from the final Transformer layer of the underlying bi-encoder;
(2) we conduct an \textit{experimental analysis} on three different DRMs\footnote{Reproducible code: \url{https://anonymous.4open.science/r/DenseRetrievalMoE}} (Contriever, BERT, and TinyBERT) investigating \texttt{SB-MoE}'s retrieval performance, as well as the impact of its hyper-parameters (i.e., number of experts used during training and inference), compared to fully fine-tuned models on four different benchmarks.

\section{Related Work}\label{sec:relatedwork}
DRMs often outperform lexicon-based models (e.g., BM25 \cite{robertson1995okapi}), since they can capture the semantic context of queries and documents by projecting them in a shared dense vector space and leverage similarity functions to score the documents according to an input query \cite{gao-callan-2022-unsupervised,kamalloo2024resources,yu-etal-2022-coco}.
We leverage three DRMs, namely Contriever \cite{izacard2022unsupervised}, BERT \cite{devlin-etal-2019-bert}, and TinyBERT \cite{jiao-etal-2020-tinybert}. 
Contriever is a state-of-the-art BERT-based model that exploits \textit{contrastive learning}, a Machine Learning technique that uses pairs of positive and negative examples to learn meaningful and distinctive representations of queries and documents. 
TinyBERT leverages \textit{knowledge distillation} \cite{romero2014fitnets} to transfer knowledge from its larger counterpart, BERT, to a tinier version, reducing training times and computational expenses. 

A common downside of DRMs is their continuous adaptation needs, which often leads to low generalizability and overall robustness \cite{liu2024robust,sidiropoulos2022analysing}.
To that regard, numerous MoE \cite{jacobs1991adaptive} approaches have been proposed in the literature, as the MoE framework can handle multiple types of data and tasks \cite{collobert2001parallel,li-etal-2021-multi-task-dense} and has been successfully used in different scenarios, such as classification tasks \cite{eigen2013learning}, and multi-lingual machine translation \cite{shazeer2017outrageously}.
In IR, MoE has been employed for tasks such as first-stage retrieval \cite{guo2024came}, passage retrieval \cite{ma2023cot}, and Q\&A \cite{Dai2022MixtureOE,shen2024mixtureofexperts}, where the feed-forward block of the underlying Transformer-based model's layers is replaced by multiple FFNs, each acting as a distinct expert.
While these approaches benefit the underlying model, they substantially increase the overall number of parameters.
To address this, DESIRE-ME\textcolor{black}{, proposed in} \cite{kasela2024desire-me}, significantly reduces the additional parameters by applying the MoE block solely on the query embedding representation outputted by the underlying DRM.
In our work, we apply the MoE block on both query and document representations and train the obtained architecture end-to-end for retrieval.

\section{Methodology}\label{sec:method}

\texttt{SB-MoE} builds upon a bi-encoder DRM architecture \cite{reimers-gurevych-2019-sentence}, which allows for independent encoding of documents and queries to enhance scalability and to enable the computation of relevance scores through a similarity function (e.g., cosine similarity). 
While it is possible to employ separate encoders \cite{karpukhin2020dense}, using a single encoder for both queries and documents improves robustness, without significantly affecting performance \cite{izacard2022unsupervised,reimers-gurevych-2019-sentence}.

\begin{figure}[t]
    \centering
    \includegraphics[width=1\linewidth]{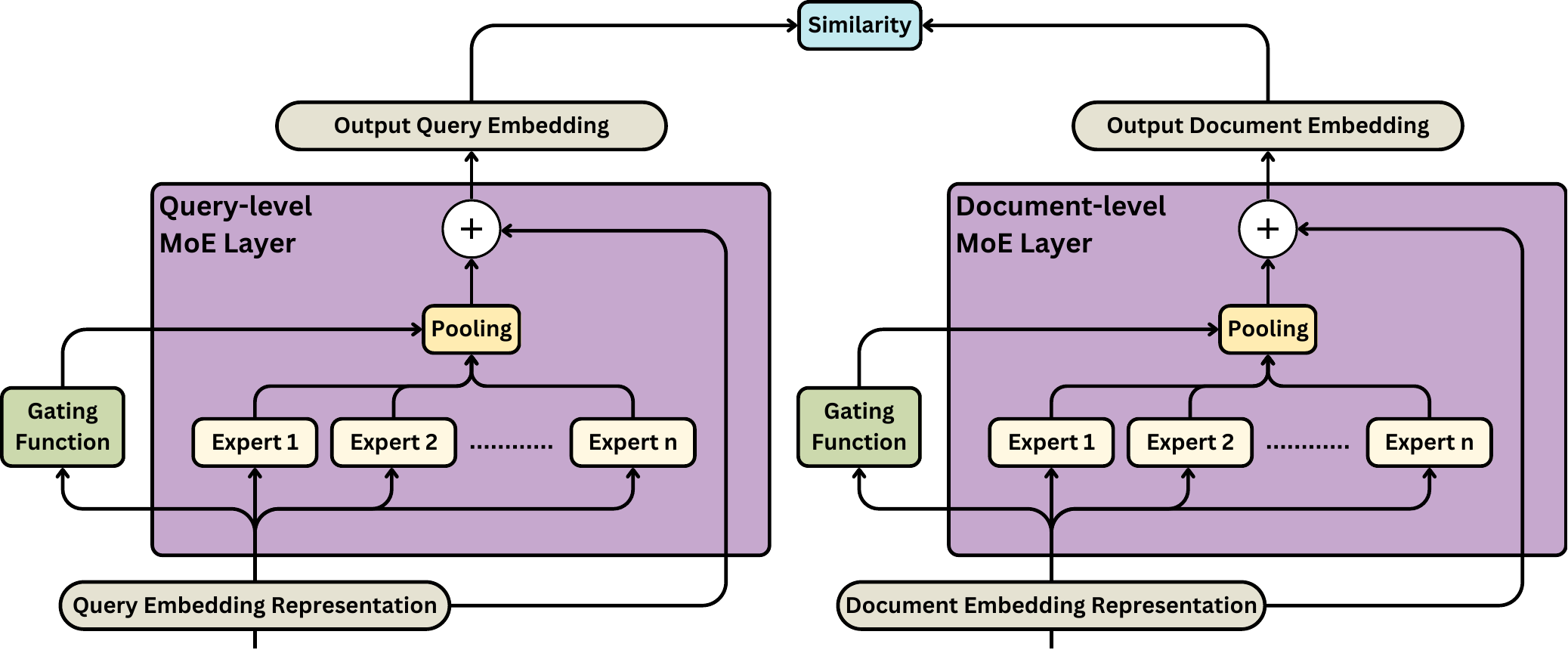}
    \caption{Overview of the \texttt{SB-MoE} architecture, highlighting its three main parts.}
    \label{fig:moe-arch}
\end{figure}

Figure \ref{fig:moe-arch} presents the \texttt{SB-MoE} architecture, which consists of three main parts: 
(1) the query and document level employed \textit{experts}, which are added after the output of the final Transformer layer of the underlying bi-encoder; 
(2) \textit{the gating function}, which is trained in an unsupervised manner to combine the experts' output for a given input; and 
(3) \textit{the pooling module} used in the final stage to aggregate the experts' representations and produce the final embedding to be used to measure similarity between the query and documents.

\textit{The experts} receive the input embedding directly from the underlying model (Fig. \ref{fig:moe-arch}) and apply a series of transformations.
The output is $n$ modified representations, where $n$ is the total number of experts.
\textit{The gating function} determines the importance of each expert’s contribution to the input query or document embedding.
It produces an $n$-dimensional vector of weights, which depicts the likelihood of each expert being suited to handle the given input.
We rely on noisy Top-1 gating, as proposed by Shazeer et al. \cite{shazeer2017outrageously}, for training the gating function.
This approach ensures that \texttt{SB-MoE} can explore different experts during training, which enhances the overall robustness of the model.
Regarding \textit{the pooling module}, we experiment with two different pooling strategies. 
The first one is Top-1 gating \cite{DBLP:conf/nips/ZhouLLDHZDCLL22} (\texttt{SB-MoE}$_{\text{TOP-1}}$) which selects solely the output of the expert that the gating function assigned the highest score to.
The second approach (\texttt{SB-MoE}$_{\text{ALL}}$) calculates probability scores from the gating function's output vector through a softmax normalization \cite{JordanHierarchical1994}, and based on them it produces the final output, which is the weighted sum of all experts' outputs.

\section{Experimental Analysis}\label{sec:experiments}
This section presents the empirical evaluation conducted to answer the following research questions (RQs):

\setlength{\leftskip}{.5cm}
\noindent \textbf{RQ1}
How does \texttt{SB-MoE} compare, in terms of effectiveness, to standard model fine-tuning?\\
\noindent \textbf{RQ2}
How does the number of experts impact the retrieval effectiveness of \texttt{SB-MoE}?

\setlength{\leftskip}{0cm}

\subsection{Experimental Settings.}

\paragraph{Datasets.} In our experiments, we use four publicly available datasets: 
(i) Natural Questions (NQ) \cite{kwiatkowski2019NQ}, an open domain Q\&A dataset based on a corpus consisting of 2.6 million Wikipedia passages, with 132k training queries with 1.2 average relevance judgments per query and 3.5k test queries; 
(ii) HotpotQA \cite{yang-etal-2018-hotpotqa}, a Q\&A dataset with multihop questions that also relies on a Wikipedia-based corpus of 5.2 million documents, with 85k training queries with 2 average relevance judgments per query and 7.4k test queries; and 
(iii) the Multi-Domain Benchmark proposed by Bassani et al. \cite{bassani2022multidomain}, from which we exploit the Political Science (PS) and Computer Science (CS) domains for the task of domain-specific academic search.
Both collections consist of 4.8 million documents, with PS, containing 160k training queries with an average of 3.8 relevance judgments per query and 5.7k testing queries, while CS contains 550k training queries with an average of 3.25 relevance judgments per query and 6.5k testing queries.
We use these datasets with distinct characteristics to evaluate \texttt{SB-MoE} under different conditions and study its consistency across various training sample sizes and tasks.

\paragraph{\texttt{SB-MoE} and Training Hyper-parameters.} 
For RQ1 we employ 6 distinct experts for all models and datasets, and for RQ2 we vary the number of experts in the range of 3-12.
Our selection is based on previous studies  
\textcolor{black}{on} the impact of the number of experts on model performance \cite{DBLP:conf/coling/LiHWYXJLLZ24,DBLP:conf/iclr/ZadouriUAELH24}, 
\textcolor{black}{which} reveal that a high number of experts does not necessarily lead to performance gains \cite{DBLP:conf/iclr/ChenZJLW23}, and set the number of experts in the range of 2-8 \cite{Dai2022MixtureOE,ma2023cot,wang-etal-2022-adamix}.
Following the architecture proposed by Houlsby et al. \cite{HoulsbyParameter2019}, each expert consists of a down-projection layer using an FFN that reduces the input dimension by half, while the output layer is an up-projection FFN layer, which restores the vector dimension to match that of the input's embedding.
A skip connection is also introduced within the MoE architecture.
The gating function consists of a single hidden layer, which reduces the vector dimension by half, and an output layer, which is an FFN with the same size as the number of experts. 
We set the training batch size to 64 and the learning rate to $10^{-6}$ and $10^{-4}$ for the underlying model and the experts respectively. 
We train TinyBERT for 30 epochs across all datasets as its smaller size allows for faster training times. 
BERT and Contriever are trained for 20 epochs due to resource constraints and longer training times, on all datasets except CS, where they are trained for 10 epochs since the collection's training queries are $\sim $3.5 times more than the second largest collection used (PS).
We use 5\% of the training sets for validation and keep only the checkpoint with the lowest validation loss.
We set the random seed to 42 and use contrastive loss \cite{izacard2022unsupervised} with a temperature of 0.05.

\paragraph{Metrics and Baselines.} 
We evaluate the experimental results using NDCG@10 and R@100, two metrics commonly used on BEIR, for comparability.
Statistical significance has been evaluated based on a two-sided paired Student's $t$-tests with Bonferroni multiple testing correction, at significance levels 0.05.
We integrate \texttt{SB-MoE} within three different DRMs\footnote{Available on HuggingFace: \href{https://huggingface.co/facebook/contriever}{Contriever}, \href{https://huggingface.co/google-bert/bert-base-uncased}{BERT}, and \href{https://huggingface.co/huawei-noah/TinyBERT_General_4L_312D}{TinyBERT}}.
We compare its retrieval effectiveness to that achieved by the underlying DRM, fine-tuned on the same training data and hyper-parameters.
We refer to these baseline experiments as \textit{Fine-tuned}.

\begin{table}[!t]
\caption{Results on all four datasets. Symbol * indicates a statistically significant difference over Fine-tuned. The best results for each model are \underline{underlined}.\\}
\label{tab:rq1}
\centering
\renewcommand{\arraystretch}{1.3}
\resizebox{1\linewidth}{!}{%
\begin{tabular}{l|l|cc|cc|cc|cc}
\multicolumn{2}{c|}{}                                  & \multicolumn{2}{c|}{NQ}              & \multicolumn{2}{c|}{HotpotQA}        & \multicolumn{2}{c|}{PS}              & \multicolumn{2}{c}{CS}               \\ \hline
Retriever                   & Variant    & NDCG@10 & R@100 & NDCG@10 & R@100 & NDCG@10 & R@100 & NDCG@10 & R@100 \\ \hline
\multirow{3}{*}{TinyBERT}   & \texttt{Fine-tuned} & .216    & .689  & .158    & .394  & .125    & .262  & .150    & .308  \\ 
                            & \texttt{SB-MoE}$_{\text{TOP-1}}$        & \underline{.219}    & .693  & .162*    & .399*  & \underline{.130*}    & \underline{.271*}  & \underline{.153*}    & .313*  \\
                            & \texttt{SB-MoE}$_{\text{ALL}}$        & .217    & \underline{.697*}  & \underline{.171*}    & \underline{.411*}  & .129*    & .270*  & \underline{.153*}    & \underline{.315*}  \\ \hline
\multirow{3}{*}{BERT}       & \texttt{Fine-tuned} & \underline{.265}    & \underline{.846}  & \underline{.372}    & \underline{.660}  & .183    & .374  & .172    & .362  \\ 
                            & \texttt{SB-MoE}$_{\text{TOP-1}}$        & .261    & .842  & .349*    & .642*  & .183    & .377*  &  \underline{.175*}        & \underline{.364}   \\ 
                            & \texttt{SB-MoE}$_{\text{ALL}}$        & .258*    & .840  & .362*    & .649*  & \underline{.184}    & \underline{.378*}  & .167*        & .355*   \\ \hline
\multirow{3}{*}{Contriever} & \texttt{Fine-tuned} & \underline{.426}    & \underline{.934}  & \underline{.672}    & \underline{.862}  & \underline{.251}    & \underline{.483}  & \underline{.224}    & .437  \\ 
                            & \texttt{SB-MoE}$_{\text{TOP-1}}$        & .416*    & .930*  & .653*    & .853*  & .250    & .479*  & .222*        & .435      \\ 
                            & \texttt{SB-MoE}$_{\text{ALL}}$        & .416*    & .932  & .667*    & .861  & \underline{.251}    & .483  & .223        & \underline{.438}      \\ 
\end{tabular}%
}
\end{table}

\subsection{Results and Discussion}

\paragraph{RQ1.}
As presented in Table \ref{tab:rq1}, \texttt{SB-MoE} exhibited a noticeable improvement in terms of NDCG@10 and Recall@100, especially with models having fewer parameters. 
For instance, on TinyBERT, \texttt{SB-MoE} leads to consistent performance gains in both metrics across all datasets, with a marked increase in HotpotQA, where \texttt{SB-MoE}$_{\text{ALL}}$ achieved an NDCG@10 score of .171 compared to .158 of the fine-tuned version.
However, for larger models like BERT and Contriever, the integration of \texttt{SB-MoE} had a marginal impact. 
For instance, on HotpotQA with BERT, \texttt{SB-MoE} achieved similar or slightly worse retrieval performance compared to \textit{Fine-tuned}.
These results suggest that in models already equipped with a substantial number of parameters, \texttt{SB-MoE}'s advantages may not be so prominent, potentially due to redundancy when additional experts are employed.
Therefore, the integration of \texttt{SB-MoE} particularly benefits lightweight models.

\begin{figure}[t]
    \centering
    \begin{subfigure}[b]{0.49\textwidth}
        \includegraphics[width=\textwidth]{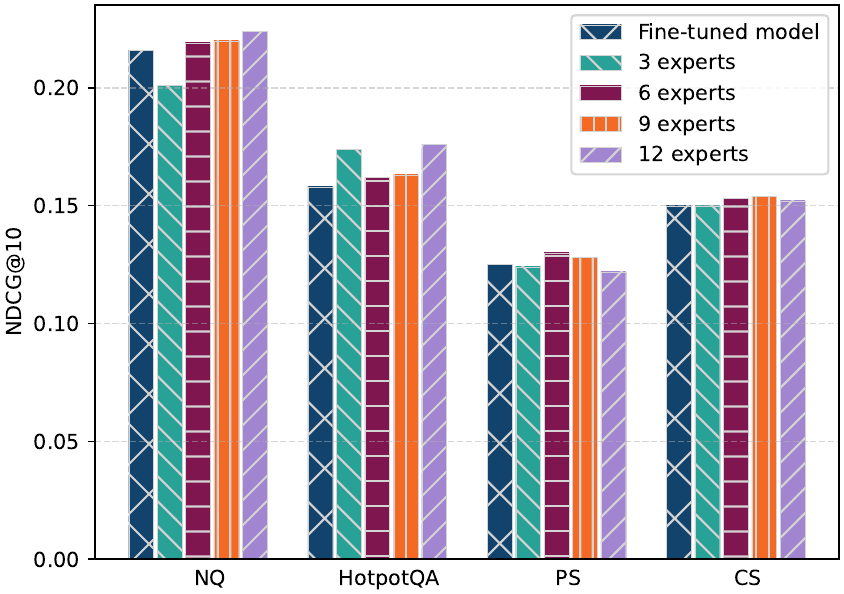}
    \end{subfigure}
    \hfill
    \begin{subfigure}[b]{0.4823\textwidth}
        \includegraphics[width=\textwidth]{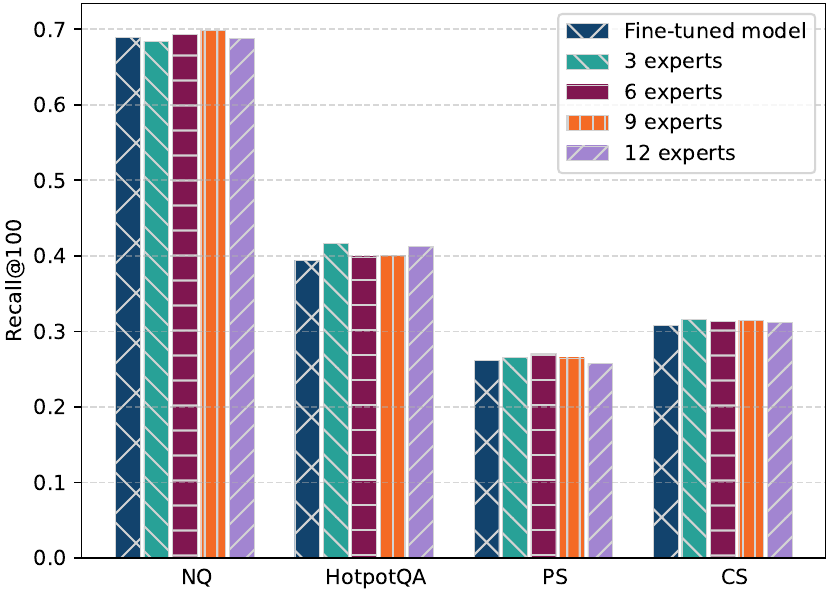}
    \end{subfigure}
    \caption{\texttt{SB-MoE}$_{\text{TOP-1}}$ on TinyBERT with 3, 6, 9, and 12 experts.}
    \label{fig:rq2}
\end{figure}

\paragraph{RQ2.}
As \texttt{SB-MoE} seems to benefit significantly lightweight models, we leverage TinyBERT to understand the impact of the number of experts, by configuring \texttt{SB-MoE} with 3, 6, 9, and 12 experts and evaluating across all datasets. 
This analysis is presented in Figure \ref{fig:rq2}.
Our investigation revealed that performance gains with different numbers of experts vary depending on the dataset, indicating that optimization is collection-dependent.
Moreover, variations in the number of experts of \texttt{SB-MoE} can lead to the maximization of different performance measures, as observed in the case of NQ, where the employment of 12 experts maximizes NDCG@10, but Recall@100 is maximized with 9 experts.
Therefore, our findings indicate that the number of employed experts is a hyper-parameter that requires tuning with respect to the domain and the addressed retrieval task.

\section{Conclusions}\label{sec:conclude}

In this work, we conduct an experimental investigation on the effectiveness of integrating a single Mixture-of-Experts block (\texttt{SB-MoE}) into Dense Retrieval Models (DRMs) to examine its potential in neural IR. 
Results show that \texttt{SB-MoE} significantly enhances retrieval performance for smaller DRMs, consistently improving NDCG@10 and R@100 across datasets.
However, larger DRMs obtained only marginal gains from \texttt{SB-MoE}, suggesting that additional experts offer limited value in models with a higher number of parameters and require dataset-specific optimization to see measurable gains.
Our analysis reveals that the number of employed experts is a hyper-parameter with a great impact on \texttt{SB-MoE}'s performance, highlighting the need for its task and domain-specific tuning.

\begin{credits}
\subsubsection{\ackname} This work has received funding from the European Union’s Horizon Europe research and innovation programme under the Marie Skłodowska-Curie grant agreement No 101073307.

\subsubsection{\discintname}
The authors have no competing interests to declare that are relevant to the content of this article.

\end{credits}

\bibliographystyle{splncs04}
\bibliography{mybibliography}

\end{document}